\documentclass[10pt, conference, letterpaper]{IEEEtran}
\IEEEoverridecommandlockouts
\usepackage{cite}
\usepackage{amsmath,amssymb,amsfonts}
\usepackage{algorithmic}
\usepackage{graphicx}
\usepackage{textcomp}
\usepackage{listings}
\usepackage{xcolor}
\usepackage{braket}
\usepackage[subrefformat=parens]{subcaption}
\usepackage{makecell}
\usepackage{hyperref}
\definecolor{terminalbg}{RGB}{30, 30, 30}
\definecolor{monokaibg}{RGB}{39, 40, 34}
\definecolor{monokaifg}{RGB}{248, 248, 242}
\lstdefinestyle{console}{
    backgroundcolor=\color{black!3},
    basicstyle=\ttfamily\scriptsize\linespread{1.2}\selectfont,
    breaklines=true,
    breakautoindent=false,
    columns=fullflexible,
    keepspaces=true,
}
\begin{document}
\bstctlcite{IEEEexample:BSTcontrol}

\title{A Model Context Protocol Server for Quantum Execution in Hybrid Quantum-HPC Environments\\
\thanks{
This work was partly performed for Council for Science, Technology and Innovation (CSTI), Cross-ministerial Strategic Innovation Promotion Program (SIP), ``Promoting the application of advanced quantum technology platforms to social issues''(Funding agency : QST).
Part of the results of this research were obtained with support from the “NEDO Challenge, Quantum Computing ``Solve Social Issues !'' by New Energy and Industrial Technology Development Organization (NEDO).
}
}

\author{\IEEEauthorblockN{Masaki Shiraishi}
\IEEEauthorblockA{\textit{Department of Applied Physics} \\
\textit{Waseda University}\\
Tokyo, Japan \\
0009-0006-9448-0181}
\and
\IEEEauthorblockN{Ikko Hamamura}
\IEEEauthorblockA{%
\textit{NVIDIA Corporation}\\
Tokyo, Japan \\
0000-0001-6416-5515}
\and
\IEEEauthorblockN{Tatsuya Ishigaki}
\IEEEauthorblockA{\textit{AIRC, AIST} \\
Tokyo, Japan \\
0000-0003-3278-234l}
\and
\IEEEauthorblockN{Tadashi Kadowaki}
\IEEEauthorblockA{\textit{G-QuAT, AIST} \\
Ibaraki, Japan}
\IEEEauthorblockA{\textit{DENSO CORPORATION} \\
Tokyo, Japan \\
0000-0002-5932-5454}
}

\maketitle

\begin{abstract}
The integration of large language models (LLMs) into scientific research is accelerating the realization of autonomous ``AI Scientists.''
While recent advancements have empowered AI to formulate hypotheses and design experiments, a critical gap remains in the execution of these tasks, particularly in the domain of quantum computing (QC).
Executing quantum algorithms requires not only generating code but also managing complex computational resources such as QPUs and high-performance computing (HPC) clusters.
In this paper, we propose an AI-driven framework specifically designed to bridge this execution gap through the implementation of a Model Context Protocol (MCP) server.
Our system enables an LLM agent to process natural language prompts submitted as part of a job, autonomously executing quantum computing workflows by invoking our tools via the MCP.
We demonstrate the framework's capability by performing essential quantum algorithmic primitives, including sampling and computation of expectation values.
Key technical contributions include the development of an MCP server for quantum execution, a pipeline for interpreting OpenQASM code, an automated workflow with CUDA-Q for the ABCI-Q hybrid platform, and an asynchronous execution pipeline for remote quantum hardware using the Quantinuum emulator via CUDA-Q.
This work validates that AI agents can effectively abstract the complexities of hardware interaction through an MCP-based architecture, thereby facilitating the automation of practical quantum research.
\end{abstract}


\section{Introduction}

Artificial Intelligence (AI) has become indispensable tools in modern scientific discovery~\cite{wang2023scientific}.
In the field of Quantum Computing (QC), specifically, AI is significantly accelerating research and development~\cite{alexeev_artificial_2025},
including decoder for quantum error correction~\cite{bausch2024learning}, quantum algorithms~\cite{nakaji2025generative}, and calibration~\cite{cao2025}.
Along with these advancements, new benchmarks are being developed to measure how well AI models perform in the quantum domain~\cite{mikuriya2025qcoderbenchmarkbridginglanguage, minami2025quantumbenchbenchmarkquantumproblem}.
These efforts help create a foundation for building more autonomous research systems.

Moving beyond isolated task assistance, the concept of an ``AI Scientist'' seeks to automate the end-to-end scientific process~\cite{aiscientist_v2}.
While the reasoning capabilities of Large Language Models (LLMs) enable autonomous hypothesis formulation and experimental design, a truly autonomous system requires more than text or code generation.
The discovery cycle is only complete when these experiments are not only translated into executable code but also managed and orchestrated across hardware or simulators, ensuring a transition from reasoning to results.

Therefore, this work specifically focuses on the execution phase.

In this paper, we propose a framework that integrates an LLM agent into the quantum algorithm development workflow.
The envisioned usage is as follows: a user submits a natural language prompt describing the desired quantum task (e.g., ``Prepare a 3-qubit GHZ state and measure it 1000 times'').
The LLM agent interprets this request, autonomously generates the corresponding OpenQASM~\cite{openqasm} circuit, selects an appropriate execution backend (GPU simulator or quantum hardware), and manages the job submission process.
The user then receives the execution results, such as measurement statistics or expectation values, without having to write low-level code or manage HPC job scripts manually.

The main contributions of this paper are summarized as follows:
    (1) Development of an MCP server that provides a tool interface for AI agents to execute quantum computing tasks. This architecture enables LLMs to handle complex hardware interactions, such as job submissions and API calls, through a standardized protocol.
    (2) Automated workflow for ABCI-Q Supercomputer~\cite{abci_q}: 
    The framework automates the generation and submission of batch job scripts for the job scheduler on ABCI-Q, thereby significantly lowering the barrier to entry for utilizing large-scale computational resources for quantum circuit simulations.
    (3) Asynchronous remote execution via CUDA-Q~\cite{cudaq}: We established an asynchronous execution pipeline for remote quantum resources, validated using the Quantinuum H2-1E emulator~\cite{quantinuum_h2_1e}.
    This module manages remote task lifecycles, enabling asynchronous, queue-based execution without blocking the agent's inference.
    (4) Demonstration of quantum algorithmic primitives: We verified the framework's utility by executing fundamental subroutines, specifically quantum measurement sampling and expectation value computation.
    These demonstrations confirm that the framework correctly translates high-level prompts into valid OpenQASM representations and retrieves accurate results.

\section{Proposed Framework and Implementation}

We present an end-to-end framework that enables an AI agent to autonomously execute quantum computing tasks within an HPC environment.
This section first provides an overview of the complete workflow.
Fig.~\ref{fig:architecture} illustrates the overall architecture of our framework.
The workflow proceeds as follows:
(1) A user submits a natural language prompt via shell script on the login node.
(2) The script queues an LLM job through the PBS scheduler.
(3) On the allocated compute node, a locally deployed LLM interprets the prompts and determines which quantum execution tool to invoke.
(4) The LLM sends tool requests to the Quantum MCP Server.
(5) The MCP server translates these requests into either a PBS batch job for local GPU simulation via CUDA-Q, or an API call to remote quantum hardware via Quantinuum REST API.
(6) Execution results are returned back to the user.

The following subsections describe each component: the two-stage PBS execution model (Section~\ref{sec:exec_arch}), the local LLM deployment (Section~\ref{sec:llm}), the MCP integration layer (Section~\ref{sec:mcp}), the ABCI-Q platform (Section~\ref{sec:abciq}), and the quantum execution backends (Sections~\ref{sec:primitives} and \ref{sec:quantum_platform}).

\subsection{Execution Architecture}
\label{sec:exec_arch}

\begin{figure*}[htbp]
    \centering
    \centerline{\includegraphics[width=1.0\linewidth]{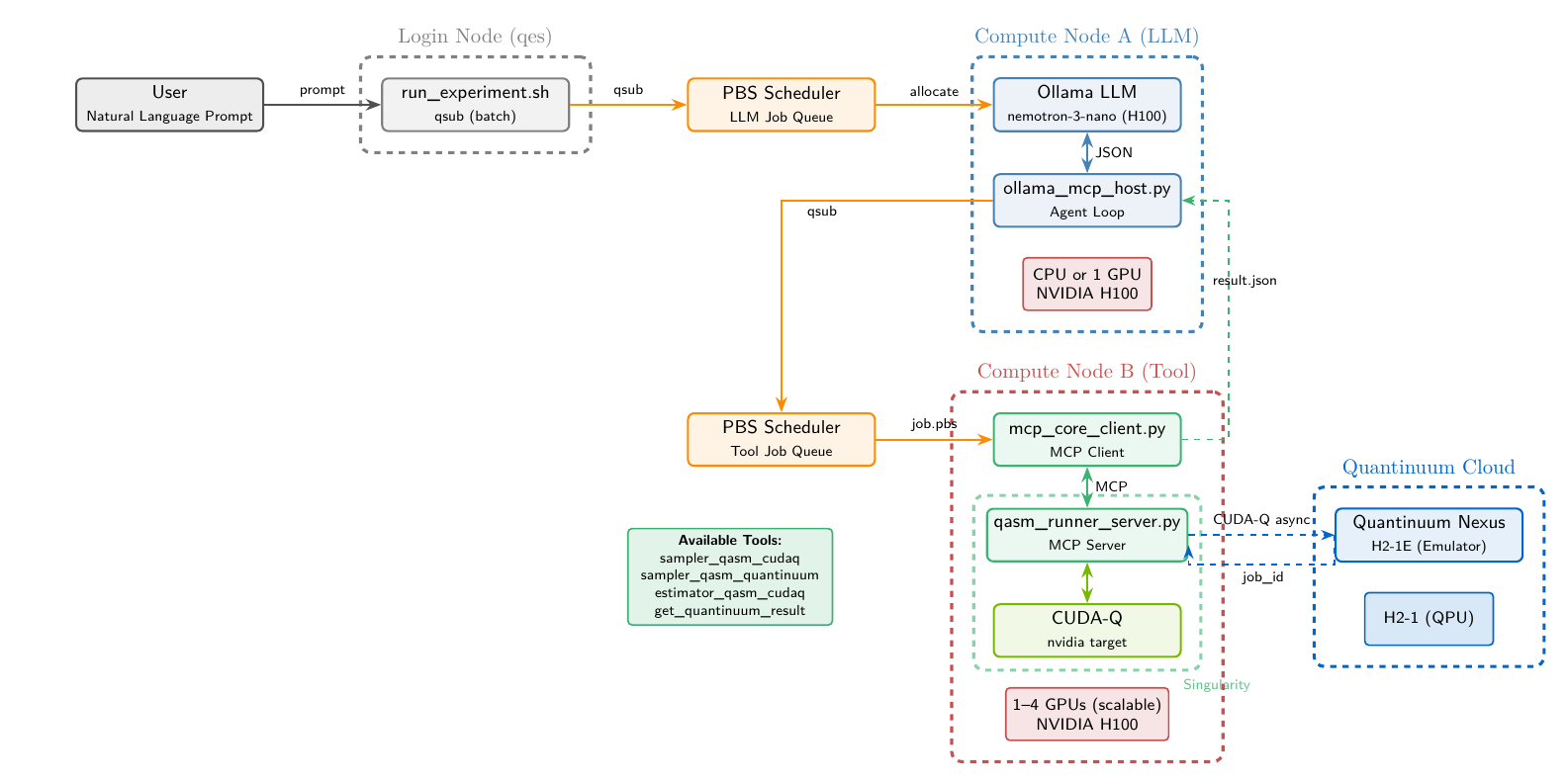}}
    \caption{%
    Two-stage PBS batch architecture. The LLM job runs on one compute node; quantum execution is dispatched as separate PBS job to a CUDA-Q node or Quantinuum cloud.
    }
    \label{fig:architecture}
\end{figure*}
Our system employs a two-stage PBS batch job architecture to avoid heavy computation on shared login nodes (Fig.~\ref{fig:architecture}):

\begin{enumerate}
    \item \textbf{LLM Job}: The user submits a prompt via \texttt{run\_experiment.sh}, which queues an LLM job through PBS.
    On the allocated compute node, the Ollama LLM server starts, and the MCP Host agent processes the prompt.
    \item \textbf{Tool Execution}: When the agent decides to execute a quantum circuit, it submits a separate PBS batch job (\texttt{qsub}) to another compute node.
    The batch job runs CUDA-Q simulation inside a Singularity container with 1--4 GPUs, or submits to Quantinuum cloud via REST API.
\end{enumerate}

This architecture separates the LLM inference from computationally intensive quantum simulations, allowing independent resource scaling.

\subsection{Local Large Language Model}
\label{sec:llm}

A design principle of our framework is to ensure that all data flows and computational processes remain strictly confined within the ABCI-Q environment.
To protect sensitive scientific data and unpublished algorithms, we adopted a local LLM approach, eliminating dependencies on external commercial cloud services and establishing a closed-loop autonomous system within the HPC infrastructure.

To realize this secure, self-contained environment, we selected NVIDIA Nemotron 3 Nano~\cite{nvidia2025nemotron3nanoopen} (31.6B parameters) as the inference engine.
By leveraging its Mixture of Experts (MoE) architecture, the model limits the number of active parameters during inference to approximately 3.6 billion.
This characteristic allows the system to maintain the high-level reasoning capabilities of a large-scale model while significantly reducing computational overhead.

The model is deployed using the Ollama runtime with 4-bit quantization,
enabling low latency operation on resource-constrained environments, such as the CPU-only nodes available in ABCI-Q.
While computationally intensive quantum circuit simulations via CUDA-Q are offloaded to high-performance NVIDIA H100 nodes via batch jobs, the agent itself is hosted locally in a secure and resource-efficient manner.
The design ensures both data sovereignty and system performance.

To enable reliable tool-calling behavior, we employed a minimal system prompt design. Rather than providing extensive instructions, the prompt contains only essential notes: OpenQASM~2.0 syntax rules, the mathematical definitions of quantum gates (e.g., $R_Z(t) = e^{-itZ/2}$, the $\mathrm{CX}$-$R_Z$-$\mathrm{CX}$ decomposition for $ZZ$ interactions), and the format for specifying observables. The prompt also defines a structured JSON output format for tool invocation (\texttt{call\_tool} / \texttt{final}) and includes the list of available tools with their parameter schemas, which are dynamically generated from the MCP server's tool definitions. This approach leverages the model's built-in reasoning through an extended thinking mode, where the model generates an internal chain-of-thought before producing its structured JSON output.

\subsection{System Integration via Model Context Protocol}
\label{sec:mcp}

To seamlessly connect the local inference engine described in Section \ref{sec:llm} with the quantum execution backend defined in Section~\ref{sec:quantum_platform}, we adopted the Model Context Protocol (MCP) \cite{anthropic2024mcp} as the integration layer. MCP is an open standard designed to bridge AI models with external data and tools, providing a universal interface for LLMs to interact with their surrounding environment.

In our framework, we encapsulated the quantum execution environment within a ``Quantum Execution MCP Server.'' 
This server exposes the execution primitives defined in Section \ref{sec:primitives} as standardized ``Tools'' discoverable by the AI agent.
This architecture allows the agent to transparently access HPC resources on ABCI-Q and control quantum hardware through a unified protocol, effectively treating complex remote job submissions as simple as local function calls.

Recent developments include the Amazon Braket MCP Server~\cite{aws-mcp},
which utilizes the Model Context Protocol (MCP) server to enable AI agents to interact with cloud-based quantum services.
Our framework addresses both local HPC resources and remote quantum resources.
It automates job submission for supercomputing clusters while managing REST API calls for remote quantum hardware.
This dual capability allows AI agents to seamlessly transition between high-performance classical simulations and actual quantum hardware execution.

\subsection{ABCI-Q}
\label{sec:abciq}
ABCI-Q is a hybrid quantum-classical supercomputing platform operated by AIST in Japan.
The system provides NVIDIA H100 GPUs for high performance simulation and is planned to integrate various quantum processors units (QPUs), including Fujitsu's superconducting QPU, QuEra's neutral atom QPU, and OptQC's photonics QPU.
The ABCI-Q leverages PBS Professional for job scheduling and resource management.

At present, while these local QPU nodes are not yet available, the platform allows for network connectivity to external quantum cloud services.
In our study, we utilized this capability to access and execute tasks on the Quantinuum via CUDA-Q.

\subsection{Quantum Algorithmic Primitives}
\label{sec:primitives}

To support practical quantum workflows, our framework addresses two fundamental execution primitives essential for modern quantum algorithms.
The first is measurement sampling, which involves collapsing the quantum state in the computational basis to obtain a distribution of bitstrings.
This primitive is critical for probabilistic tasks such as combinatorial optimization and quantum state characterization.
The second is expectation value estimation, which computes the average value of a given observable $\hat{O}$ (e.g., a Hamiltonian for energy estimation) with respect to a quantum state $\ket{\psi}$, denoted as $\braket{\psi|\hat{O}|\psi}$.
This operation serves as the core subroutine for evaluating cost functions in Variational Quantum Algorithms (VQAs), including the Variational Quantum Eigensolver (VQE)~\cite{peruzzo2014variational} and the Quantum Approximate Optimization Algorithm (QAOA)~\cite{farhi2014quantum}.

\subsection{Quantum Platform}
\label{sec:quantum_platform}

To implement these execution primitives efficiently, we leverage CUDA-Q, an open-source platform designed for heterogeneous quantum-classical computing~\cite{cudaq}.
CUDA-Q provides a unified programming model that seamlessly integrates CPUs, GPUs, and QPUs.
A key advantage of CUDA-Q for our framework is its provision of high-level APIs that directly map to the aforementioned primitives.
Specifically, CUDA-Q implements \texttt{cudaq.sample} for retrieving measurement statistics (sampling) and \texttt{cudaq.observe} for computing expectation values of observables.
By utilizing these standardized functions, our framework abstracts the low-level intricacies of backend management, enabling the AI agent to dispatch tasks to various backends—ranging from GPU accelerated simulators NVIDIA cuQuantum~\cite{cuquantum2023} to quantum hardware through a consistent and unified interface.

To facilitate experimentation across diverse computing environments, our platform incorporates a unified execution layer that supports multiple backends.
For GPU-accelerated simulations on the ABCI-Q supercomputer, the platform utilizes CUDA-Q as the primary execution engine.
In addition to this local HPC integration, the platform is equipped with a dedicated module for remote execution on the Quantinuum H2-1E emulator~\cite{quantinuum_h2_1e}.
This remote module implements the necessary functionality to interface with external servers via a REST API, supporting asynchronous job submission and status polling to accommodate queue-based processing.
By providing these distinct execution paths through a single interface, the platform enables the AI agent to dispatch quantum tasks to either high-performance GPU clusters or remote quantum cloud services without requiring modifications to the core algorithm logic.

\section{Experimental Results}
We validated our framework by executing two fundamental quantum algorithmic primitives: sampling and expectation value estimation.
Experiments were conducted on ABCI-Q using NVIDIA H100 GPUs with CUDA-Q's \texttt{nvidia} target for GPU-accelerated state vector simulation, as well as on Quantinuum's H2-1E trapped-ion emulator via REST API for cloud-based quantum execution.

\subsection{Sampling: GHZ State Preparation}

To demonstrate the sampling primitive, we instructed the agent to prepare a 3-qubit Greenberger-Horne-Zeilinger (GHZ) state and measure it 2000 times.
The agent autonomously:
\begin{enumerate}
    \item Generated the OpenQASM 2.0 circuit for GHZ state preparation
    \item Selected the \texttt{sampler\_qasm\_cudaq} tool
    \item Submitted a PBS batch job for execution
    \item Returned the measurement results
\end{enumerate}

\begin{figure}[!htbp]
\centering
\begin{lstlisting}[style=console]
[<user_name>@qes04 abciq-mcp-server]$ ./run_experiment.sh --prompt "Prepare a 3-qubit GHZ state in OpenQASM 2.0 and measure it 2000 times using sample
r_qasm_cudaq."
=== Submitting LLM Job ===
Job directory: /home/<username>/abciq-mcp-server/.llm_jobs/<job_id>
PBS group: qqch50095
LLM resources: rt_QG=1, walltime=01:00:00
Tool resources: rt_QG=1, walltime=00:20:00

Submitted: 105211.qjcm

Waiting for job to complete...

=== LLM Job Started on qh002 ===
Job ID: 105211.qjcm
Date: Wed Mar  4 05:32:07 PM JST 2026

Starting Ollama server...
Ollama PID: 3289261
Ollama is ready (attempt 2)

=== Running MCP Host ===

=== Job Completed ===
Exit code: 0

Result:
{
  "openqasm_code": "#include \"qelib1.inc\";\\nqreg q[3];\\ncreg c[3];\\nh q[0];\\ncx q[0], q[1];\\ncx q[0], q[2];\\nmeasure q -> c;",
  "shots": 2000,
  "counts": {
    "000": 1040,
    "111": 960
  },
  "probabilities": {
    "000": 0.52,
    "111": 0.48
  }
}
\end{lstlisting}
\caption{%
LLM job submission to ABCI-Q.
The console illustrates the terminal where the AI agent automatically generates an OpenQASM for a GHZ state and submits it as a batch job to the ABCI-Q GPU node via the PBS Pro.
The framework manages the entire lifecycle of the task, including PBS job script generation, job submission, and retrieval of the results.
The resulting measurement distribution shows that $|000\rangle$ and $|111\rangle$ are observed with approximately equal frequency (roughly 50\% each).
}
\label{fig:GHZ}
\label{fig}
\end{figure}

The results showed $|000\rangle$: 996 counts (49.8\%) and $|111\rangle$: 1004 counts (50.2\%), consistent with the theoretical 50-50 distribution of an ideal GHZ state (Fig.~\ref{fig:GHZ}).

\subsection{Quantinuum Cloud Execution}

\begin{figure*}[ht]
\centering
\begin{minipage}{.49\linewidth}
\centering
\begin{lstlisting}[style=console]
Prepare a 3-qubit GHZ state in OpenQASM 2.0 and submit it to Quantinuum H2-1E emulator with 100 shots using sampler_qasm_quantinuum. 
\end{lstlisting}
\subcaption{GHZ experiment on Quantinuum (submission)}
\label{fig:GHZ_Quantinuum}
\end{minipage}
\begin{minipage}{.5\linewidth}
\centering
\begin{lstlisting}[style=console]
Use get_quantinuum_result to retrieve the results.
\end{lstlisting}
\subcaption{GHZ experiment on Quantinuum (results)}
\label{fig:GHZ_Quantinuum_result}
\end{minipage}
\caption{%
Asynchronous execution on the Quantinuum H2-1E emulator.
(a) The agent submits the circuit and receives a job ID.
(b) Result retrieval after remote execution completes.
}
\end{figure*}

To demonstrate remote quantum hardware integration, we extended the same GHZ experiment to Quantinuum's trapped-ion system via the REST API.
Unlike local simulation, cloud execution requires asynchronous job handling due to queue wait times.

The agent workflow for Quantinuum execution:
\begin{enumerate}
    \item Selected the \texttt{sampler\_qasm\_quantinuum} tool with target device H2-1E (emulator)
    \item Submitted the circuit via Quantinuum REST API and received a job ID
    \item Later invoked \texttt{get\_quantinuum\_result} to retrieve completed results
\end{enumerate}

The results from the H2-1E emulator with 100 shots showed $|000\rangle$: 42 counts (42\%), $|111\rangle$: 57 counts (57\%), and $|110\rangle$: 1 count (1\%) (Fig.~\ref{fig:GHZ_Quantinuum_result}).
The small deviation from ideal 50-50 distribution and the single bit-flip error ($|110\rangle$) reflect the realistic noise model of Quantinuum's trapped-ion emulator, demonstrating our framework's capability to interface with production quantum cloud services.

\subsection{Estimation: QAOA Cost Function}
To demonstrate the expectation value primitive, we tasked the agent with executing a Quantum Approximate Optimization Algorithm (QAOA)~\cite{farhi2014quantum} workflow.
QAOA is designed to find approximate solutions to combinatorial optimization problems by minimizing the expectation value of a cost Hamiltonian $H_C$, which encodes the objective function of the problem. For MaxCut, the objective is to maximize the number of edges crossing the cut, and we evaluate $\langle H_C \rangle$ for a fixed parameter instance in this validation.

The QAOA circuit is generally defined by an ansatz consisting of $p$ layers of alternating unitaries.
Each layer $k \in \{1, \dots, p\}$ comprises a cost unitary $U_C(\gamma_k) = e^{-i\gamma_k H_C}$ implemented via $ZZ$ rotations and a mixer unitary $U_M(\beta_k) = e^{-i\beta_k \sum_j X_j}$ realized through $R_X(2\beta_k)$ gates.

For this validation, we selected the Max-Cut problem on a triangle graph for the demonstration.
The cost Hamiltonian for this 3-qubit system is expressed as:
\begin{equation}
    H_C = 1.5I - 0.5(Z_0 Z_1 + Z_1 Z_2 + Z_0 Z_2).
\end{equation}

For the validation of our framework, we implemented a single-layer ansatz ($p=1$).
We applied Hadamard gates on all qubits to prepare the initial uniform superposition.
To evaluate the execution of the expectation value computation, the variational parameters for this instance were specifically bound to $\gamma = 1.4$ and $\beta = 0.8$.

The agent:
\begin{enumerate}
    \item Received the OpenQASM 2.0 circuit with parameters pre-substituted:
{\small
\begin{verbatim}
OPENQASM 2.0;
include "qelib1.inc";
qreg q[3];
h q[0]; h q[1]; h q[2];
cx q[0],q[1]; rz(-1.4) q[1]; cx q[0],q[1];
cx q[1],q[2]; rz(-1.4) q[2]; cx q[1],q[2];
cx q[0],q[2]; rz(-1.4) q[2]; cx q[0],q[2];
rx(1.6) q[0]; rx(1.6) q[1]; rx(1.6) q[2];
\end{verbatim}
}
    \item Selected the \texttt{estimator\_qasm\_cudaq} tool with observable specified as JSON:
{\small
\begin{verbatim}
observable_terms: [
  {"coeff": 1.5, "pauli": ""},
  {"coeff": -0.5, "pauli": "Z0 Z1"},
  {"coeff": -0.5, "pauli": "Z1 Z2"},
  {"coeff": -0.5, "pauli": "Z0 Z2"}
]
\end{verbatim}
}
    \item Executed analytic expectation value calculation (\texttt{shots: null}) via PBS batch job
\end{enumerate}

\begin{figure}[htbp]
\centering
\begin{lstlisting}[style=console]
[<user_name>@qes04 abciq-mcp-server]$ ./run_experiment.sh --prompt "Compute the expectation value of the MaxCut Hamiltonian for a triangle graph using
 a 3-qubit QAOA. ansatz (p=1, gamma=1.4, beta=0.8). The cost Hamiltonian is: H = 1.5*I - 0.5*(Z0Z1 + Z1Z2 + Z0Z2). Use estimator_qasm_cudaq with the
 appropriate OpenQASM 2.0 circuit and observable_terms."
=== Submitting LLM Job ===
Job directory: /home/<user_name>/abciq-mcp-server/.llm_jobs/<job_id>
PBS group: qqch50095
LLM resources: rt_QG=1, walltime=01:00:00
Tool resources: rt_QG=1, walltime=00:20:00

Submitted: 105242.qjcm

Waiting for job to complete...

=== LLM Job Started on qh453 ===
Job ID: 105242.qjcm
Date: Wed Mar  4 05:45:06 PM JST 2026

Starting Ollama server...
Ollama PID: 799753
Ollama is ready (attempt 2)

=== Running MCP Host ===

=== Job Completed ===
Exit code: 0

Result:
{
  "expectation": 0.029909259639680386
}
\end{lstlisting}
\caption{%
Demonstration of QAOA expectation value computation on ABCI-Q GPU nodes.
The console displays the execution of an expectation value computation for QAOA with the MCP server.
The output illustrates the computed expectation value of the Hamiltonian, $\braket{ \psi(\beta, \gamma) | H_C | \psi(\beta, \gamma) }$.
}
\label{fig:QAOA}
\end{figure}

The computed expectation value was $\langle H_C \rangle = 0.0299$ (Fig.~\ref{fig:QAOA}), which matches the theoretical value obtained from direct numerical simulation of the quantum circuit.
This result demonstrates the agent's capability to handle multi-term observables and parameterized quantum circuits, which are essential components for variational quantum algorithms.
\subsection{Quantitative Evaluation}

To assess the reliability of the optimized system prompt, we
conducted multiple independent runs for each task.
The QAOA task achieved a 100\% success rate over five runs
(average 4 agent steps), consistently producing the
theoretically expected expectation value of $0.0299$ with
correct gate.
The GHZ task succeeded in six of eight runs (average 3 steps),
with all runs producing correct quantum circuits; the two
failures were caused by malformed JSON output formatting
rather than incorrect quantum logic.

\begin{table}[t]
\centering
\caption{Average end-to-end latency breakdown.
LLM: all inference calls with extended thinking;
PBS Sched.: job queue wait and container startup;
Exec: quantum computation (CUDA-Q simulation or
Quantinuum);
OH: NFS synchronisation waits, polling granularity,
and \texttt{qsub} submission latency.
CUDA-Q rows are averaged over five runs;
the Quantinuum row over four runs.}
\label{tab:latency}
\begin{tabular}{lrrrrr}
\hline
Task & LLM (s) & \shortstack{PBS\\Sched.\ (s)} & Exec (s) & OH (s) & Total (s) \\
\hline
QAOA          & 43.6 & 57.1 & 1.0 &  5.6 & 107.3 \\
GHZ           & 18.0 & 94.0 & 0.5 & 12.1 & 124.6 \\
\shortstack[l]{GHZ\\(Quantinuum)} & 35.6 & 131.0 & 8.3 & 16.9 & 191.8 \\
\hline
\end{tabular}
\end{table}

We performed a benchmarking analysis of job execution by LLMs (Table~\ref{tab:latency}).
Across all tasks, quantum computation (Exec column)
is negligible: under 1\% for CUDA-Q and 4.3\% for
Quantinuum. The dominant bottleneck is PBS job scheduling,
which incurs an irreducible floor of $\sim$42\,s per job
for container startup and runtime initialization.
The Quantinuum task requires three PBS jobs per run
(error retry, circuit submission, and result retrieval),
accounting for its longer total time; the cloud API
interaction itself takes only 8.3\,s.

\section{Conclusion}

In this work, we addressed the critical bottleneck of ``execution'' in the automation of scientific discovery by developing a specialized Quantum Execution MCP Server capable of autonomously handling quantum computing tasks.
By integrating local LLMs with CUDA-Q via the Model Context Protocol (MCP) within the ABCI-Q supercomputing environment, we established a unified workflow spanning from quantum circuit simulation to execution on quantum hardware emulator on the quantum cloud service.

The contributions of this framework extend beyond simple automation.
First, by providing a robust execution layer, our system acts as the ``hands'' for existing reasoning agents, enabling the autonomous execution of complex quantum experiments.
Second, it democratizes access to high-performance computing by abstracting infrastructure complexities, empowering quantum physicists to utilize supercomputing power without deep HPC expertise.
Third, our architecture, which leverages local LLMs within a closed HPC network, establishes a model for data sovereignty.
This ensures that research involving sensitive algorithms or proprietary data can leverage AI capabilities without the risk of external data leakage.

Future work will focus on extending the current workflow to support interactive execution sessions.
We also aim to implement more sophisticated resource management capabilities to dynamically execute tasks across heterogeneous compute nodes (CPUs, GPUs, and QPUs) based on workload characteristics.
Furthermore, addressing scientific reproducibility remains a priority.
To mitigate the nondeterministic nature of AI generation, logging and error handling mechanisms are required.
Developing features that ensure the reproducibility of AI workflows is essential to building trust in the findings of AI scientists.

\section*{Acknowledgment}
\addcontentsline{toc}{section}{Acknowledgment}

The results presented in this paper were obtained using AIST G-QuAT's ABCI-Q.

Masaki Shiraishi was supported by an internship at Jij Inc.

\bibliographystyle{IEEEtran}
\bibliography{ref}
\end{document}